\documentclass[sigconf,authorversion]{acmart}
\usepackage{xcolor}
\definecolor{darkblue}{rgb}{0,0,0.4}

\copyrightyear{2021}
\acmYear{2021}
\setcopyright{acmlicensed}\acmConference[ARES 2021]{The 16th International Conference on Availability, Reliability and Security}{August 17--20, 2021}{Vienna, Austria}
\acmBooktitle{The 16th International Conference on Availability, Reliability and Security (ARES 2021), August 17--20, 2021, Vienna, Austria}
\acmPrice{15.00}
\acmDOI{10.1145/3465481.3470081}
\acmISBN{978-1-4503-9051-4/21/08}

\usepackage{glossaries}

\newacronym{CA}{CA}{Certification Authority}
\newacronym{CDN}{CDN}{Content Distribution Network}
\newacronym{CERT}{CERT}{Computer Emergency Response Team}
\newacronym{CSIRT}{CSIRT}{Computer Security Incidence Response Team}
\newacronym{CSP}{CSP}{Content Security Policy}
\newacronym{CSR}{CSR}{Corporate Social Responsibility}
\newacronym{CMS}{CMS}{Content Management System}
\newacronym{DDoS}{DDoS}{Distributed Denial of Service}
\newacronym{DKIM}{DKIM}{DomainKeys Identified Mail}
\newacronym{DNS}{DNS}{Domain Name System}
\newacronym{DPA}{DPA}{Data Protection Authority}
\newacronym{EU}{EU}{European Union}
\newacronym{FIRST}{FIRST}{Forum of Incident Response and Security Teams}
\newacronym{GDPR}{GDPR}{General Data Protection Regulation}
\newacronym{HSTS}{HSTS}{HTTP Strict Transport Security}
\newacronym{HTML}{HTML}{Hypertext Markup Language}
\newacronym{IA}{IA}{Institutional Approach}
\newacronym{IDS}{IDS}{Intrusion Detection System}
\newacronym{ISP}{ISP}{Internet Service Provider}
\newacronym{IoT}{IoT}{Internet of Things}
\newacronym{IXP}{IXP}{Internet eXchange Point}
\newacronym{MX}{MX}{Mail eXchange}
\newacronym{NGO}{NGO}{Non-Governmental Organization}
\newacronym{NTP}{NTP}{Network Time Protocol}
\newacronym{OS}{OS}{Operating System}
\newacronym{PCIDSS}{PCI DSS}{Payment Card Industry Data Security Standard}
\newacronym{RbV}{RbV}{Resource-based View}
\newacronym{SSH}{SSH}{Secure Shell}
\newacronym{SPF}{SPF}{Sender Policy Framework}
\newacronym{TLD}{TLD}{Top-Level Domain}
\newacronym{TLS}{TLS}{Transport Layer Security}
\newacronym{VCS}{VCS}{Version-Control System}
\newacronym{VPN}{VPN}{Virtual Private Network}
\newacronym{XSS}{XSS}{Cross-Site Scripting}

\newcommand{\ie}[0]{i.\,e.}
\newcommand{\eg}[0]{e.\,g.}

\begin{document}

\title[Best Practices for Notification Studies for Security and Privacy Issues on the Internet]{Best Practices for Notification Studies\\for Security and Privacy Issues on the Internet}

\author{Max Maass}
\email{mmaass@seemoo.tu-darmstadt.de}
\orcid{0000-0001-9346-8486}
\affiliation{%
  \institution{Technical University Darmstadt}
  \city{Darmstadt}
  \country{Germany}
}
\author{Henning Pridöhl}
\email{henning.pridoehl@uni-bamberg.de}
\author{Dominik Herrmann}
\orcid{0000-0002-7374-3054}
\email{dominik.herrmann@uni-bamberg.de}
\affiliation{%
  \institution{Otto-Friedrich-Universität Bamberg}
  \city{Bamberg}
  \country{Germany}
}

\author{Matthias Hollick}
\orcid{0000-0002-9163-5989}
\email{mhollick@seemoo.tu-darmstadt.de}
\affiliation{%
  \institution{Technical University Darmstadt}
  \city{Darmstadt}
  \country{Germany}
}

%
%
%
%
%


\begin{abstract}
  Researchers help operators of vulnerable and non-compliant internet services by individually notifying them about security and privacy issues uncovered in their research.
  To improve efficiency and effectiveness of such efforts, dedicated notification studies are imperative.
  As of today, there is no comprehensive documentation of pitfalls and best practices for conducting such notification studies, which limits validity of results and impedes reproducibility.
  Drawing on our experience with such studies and guidance from related work, we present a set of guidelines and practical recommendations, including initial data collection, sending of notifications, interacting with the recipients, and publishing the results.
  We note that future studies can especially benefit from extensive planning and automation of crucial processes, i.\,e., activities that take place well before the first notifications are sent.
\end{abstract}

\maketitle

\section{Introduction}

When researchers discover new vulnerabilities or compliance violations, a large number of internet services may be affected. Large-scale network scans, for instance, for servers affected by the Heartbleed vulnerability \cite{Durumeric2014}, have shown that many service providers fail to secure their systems even when a vulnerability is widely discussed. Realizing that publicly announcing vulnerabilities is not sufficient, security researchers have begun to approach service providers individually.
As studies on notification effectiveness \cite{Vasek2012,Canali2013,Kuhrer2014,Durumeric2014,Cetin2016,Li2016Usenix,Li2016WWW,Stock2016,Cetin2017,Stock2018,Cetin2018,Zeng2019,Cetin2019,usenix2021,maass2021leaks} have returned inconclusive results in many areas, more studies will likely follow.

For this paper, we consider \textbf{notification studies} that are designed as follows. Having identified a security or privacy issue affecting a large number of services or websites on the internet, researchers first obtain a list of affected targets. For each target, they determine a way to reach a point of contact. Next, they notify the points of contact about the issue. To study notification effectiveness, researchers split the targets into different treatment groups. Then, they analyze responses and remediation tactics (by re-scanning the targets) and, optionally, the use of self-service tools mentioned in the notifications. Researchers may also ask the points of contact to participate in interviews or surveys.

So far, there is no established methodology for internet notification studies, which is problematic for two reasons. Firstly, there are numerous design decisions that cannot be amended at a later time, and there is much potential for implementation mistakes that may degrade the validity of the results (see, e.\,g., \cite{usenix2021}). Secondly, different study designs impede comparisons with results obtained in other studies. Both problems can be addressed by standardization of methods and following best practices.




To this end, we documented pitfalls and lessons learned while conducting a series of notification studies \cite{MaassWHH19,KrogerLH20,usenix2021,maass2021leaks} over the last years. The derived best practices presented in this paper also take into account guidance from related work (Sect.~\ref{sec:relatedwork}). We describe the design space of notification studies and their typical execution timeline (Sect.~\ref{sec:design}). We also review legal and ethical obligations (Sect.~\ref{sec:legalethics}). After that, we present guidance on data collection (Sect.~\ref{sec:datacollection}), notification handling (Sect.~\ref{sec:interaction}), and publication (Sect.~\ref{sec:publication}).
Data analysis methods are beyond the scope of this paper.

This paper may serve as both a blueprint and a checklist for future notification studies. The best practices may also be of interest for other kinds of studies, \eg{}, large-scale vulnerability scanning \cite{Cui2011} and experiments that involve interactions with a large number of service providers \cite{KrogerLH20}.



\section{Related Work}
\label{sec:relatedwork}

To derive a design space and best practices for notification experiments, we rely on our own experience and on previous work. So far, there are no dedicated publications on methods for notification studies. As notification studies typically rely on network scans, we review publications on best practices for empirical network research. A comprehensive resource is the guide by Bajpai \emph{et al.} \cite{BajpaiBFKPSSWW19}, which discusses best practices in conducting and documenting networking research, including measurements and human-subject studies. Cui and Stolfo \cite{Cui2011} report on their practical experience while running a large-scale vulnerability scanner and derive procedures for large-scale, secure and responsible vulnerability scans. Durumeric \emph{et al.} similarly discuss procedures to ensure good citizenship for large-scale internet scans \cite[Sect. 5]{Durumeric2013}.

Moreover, best practices for human subjects research on the internet are relevant. Notification studies may conflict with ethical and legal obligations, \eg{}, when they are designed as covert experiments with elements of deception \cite{Cetin2016,usenix2021}. While researchers typically discuss how they addressed ethical issues, legal obligations are mostly neglected. Mazel \emph{et al.} \cite{Mazel2017} found that not all existing scanning projects provide even basic documentation about their actions. Vitak \emph{et al.} \cite{Vitak2016} surveyed the ethical views of the online data research community and found heterogeneous results, showing that no consensus has been reached so far and encouraging greater discussion with colleagues on research ethics.

\section{Study Designs}
\label{sec:design}

In this section, we describe the design space of notification studies and their typical timeline. The purpose of this section is to provide an overview. Apart from some remarks, we postpone the presentation of best practices to later sections for two reasons. Firstly, many recommendations affect multiple components and span multiple phases. Secondly, some best practices are motivated by legal and ethical obligations, which we describe in the upcoming Section~\ref{sec:legalethics}.

\subsection{Design Space}

We focus on real-world field experiments, not on laboratory settings. General advice on robust study designs is given by Krol \emph{et al.} \cite{Krol16}.
The design space of such studies covers three broad areas.

\paragraph{Issue at Hand and Target}

The most obvious design decision is the \emph{considered issue} (\eg{}, a particular vulnerability in web applications) and the \emph{responsible party} (\eg{}, the owner of a website). The considered issue could be a vulnerability in a component (either already well-known or not) or a generic weakness such as an SQL injection. It could affect components on servers, network devices, or on clients that result from design, implementation, or configuration mistakes.

Note that the \emph{responsible party} (\eg{}, the operator of a vulnerable server) is not necessarily the one that is impacted by an issue, \ie{}, the costs are borne by others. An example of an issue that involves such a \emph{negative externality} is a system that can be misused to amplify the impact of denial-of-service-attacks \cite{Kuhrer2014,Li2016Usenix,Cetin2019}.

The responsible party is also not necessarily the \emph{recipient of the notifications}. Notifications could also be sent to ISPs that host a vulnerable server \cite{Vasek2012,Cetin2016,Stock2016,Cetin2017} or to a coordinating body like a CERT \cite{Kuhrer2014,Li2016Usenix,Stock2016}.
Previous studies have used many contact channels to reach the recipients, including addresses harvested from the WHOIS interface \cite{Vasek2012,Durumeric2014,Cetin2016,Li2016Usenix,Li2016WWW,Stock2016,Cetin2017,Stock2018,Zeng2019}, standard aliases \cite{Canali2013,Stock2016,Stock2018,Cetin2017}, or even manually collected address information \cite{Stock2018,usenix2021}.

\paragraph{Actual and Purported Sender}

Another factor is the \emph{sender of notifications}, which has been shown to affect remediation rates considerably in one study \cite{usenix2021} (although others found only a small impact \cite{Cetin2016,Zeng2019,Stock2018}). Senders may openly affiliate themselves with a university or act as private individuals. Senders with potentially more authority are CERTs, ISPs, and data protection authorities, although it can be difficult to gain access to them \cite{Stock2016}.

The choice of the sender is related to the \emph{covertness of the study}. To avoid biases such as the observer effect, it may be desirable to cover up the fact that an experiment takes place. Keeping recipients in the dark over longer periods of time is challenging. When operational mishaps or oversights in communication give the experiment away, recipients may change their behavior, invalidating the results. Even in the absence of errors, recipients that have received different treatments may learn about each other, for instance, on social media platforms or when their systems are being run by the same service provider \cite{usenix2021,MaassWHH19}. While deception may be necessary for ecological validity \cite{Egelman10,Krol16}, it mandates extensive ethical considerations and may conflict with data protection obligations.

\paragraph{Employed Instruments}

At least two more aspects have to be considered, the \emph{notification channel} (email, letters, phone calls, social media) and the \emph{number of notifications per recipient} (initial notification, reminders, separate debriefing). Both can have an impact on remediation rates. Studies may also employ other instruments, such as including an \emph{invitation to participate in a survey or interview} \cite{Li2016Usenix,Cetin2017,Zeng2019,usenix2021,Durumeric2014,Stock2018} or \emph{self-service tools} \cite{Zeng2019,Li2016WWW,Cetin2016,usenix2021}, \eg{}, an online tool to verify the issue independently or tutorials for remediation. These instruments involve more design decisions, \eg{}, about the actual implementation of an online tool. Operating an online tool oneself makes it more challenging to run a study covertly while recommending an existing third-party tool introduces operational risks beyond one's control.

\subsection{Study Timeline}
\label{sec:timeline}

\begin{figure*}
\centering
    \includegraphics[width=1\textwidth]{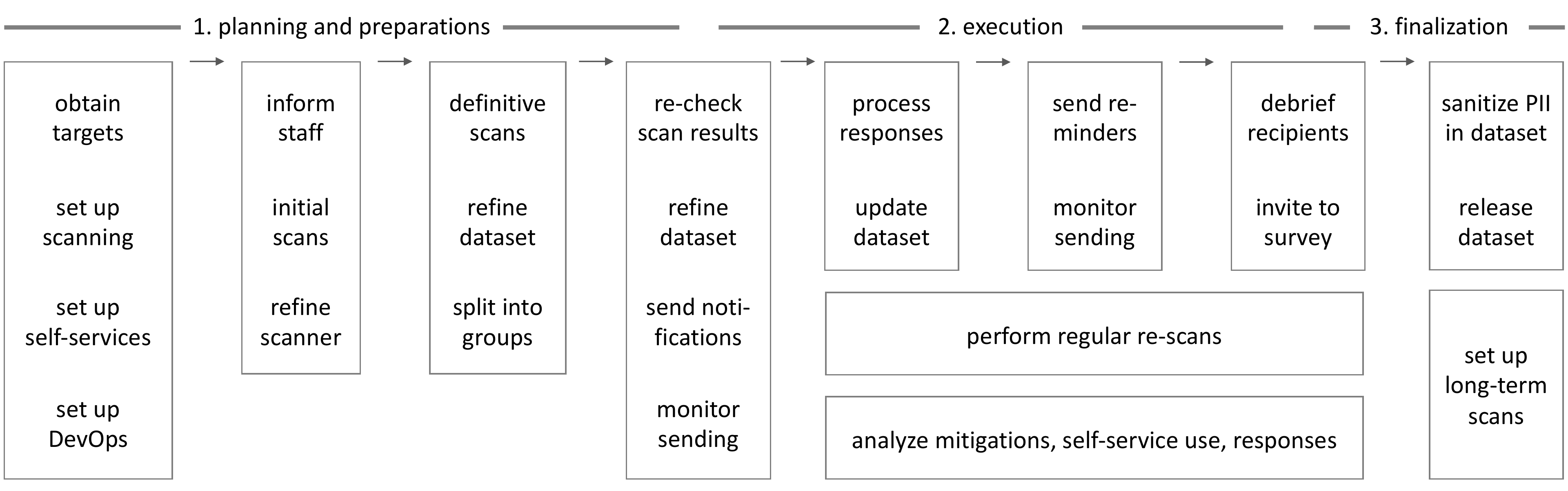}
    \caption{Idealized Timeline of a Notification Study}%
    \label{fig:timeline}%
\end{figure*}

As empirical experiments, notification studies consist of three stages: planning and preparation, execution, and finalization.
In the following, we describe the activities in these stages in chronological sequence (Fig.~\ref{fig:timeline}). For conciseness, we focus on notification studies that target \emph{service providers} rather than end users.

\paragraph{Planning}

Among the first activities are problem formulation and sketching the study design \cite{BajpaiBFKPSSWW19}.
Researchers have to determine the issue at hand, the target group for the notifications, and relevant research hypotheses.
As it is difficult to make amendments once notifications have been sent, possible reactions of recipients and treatment options should be anticipated during planning.
To this end, one should discuss the planned study design with researchers and practitioners from relevant fields, including information security, data protection, and legal scholars.
Moreover, talking to trusted members of the target group can be helpful to understand their perspective.
Finally, local Institutional Review Boards (IRB) should be approached with a comprehensive description of the study design. In our experience, clarifications of ethical considerations need ample amounts of time, also because they may necessitate refinements of the study design.

\paragraph{Infrastructure Setup}

Typically, notification studies involve scanning thousands of servers and notifying hundreds of recipients.
Scanning, data collection, and message handling should be automated with suitable tools, either off-the-shelf or custom-built.
For complex and long-running studies, we strongly recommend using development and operations (DevOps) tools such as version control systems, issue trackers, and service monitoring.
Cui and Stolfo offer further recommendations on scalability and security \cite{Cui2011}.

\paragraph{Initial Scans}

Before the first scans take place, relevant local staff (research group leaders, network administrators, people acting as abuse contacts, and senior staff in the department) should be informed about the scans.
To develop a robust scanning and data collection engine, researchers should run a series of initial scans to gain experience with error cases and refine their scanner until it works unattended and produces the same (or explainably dissimilar) results to achieve \emph{repeatability} \cite{acmartifact}.

\paragraph{Building Dataset}

The scanner is then used to collect the dataset of affected services. For these services, points of contact have to be collected.
Typically, the contact data will have to be sanitized, and the dataset may have to be refined.
Records with the same points of contact should be deduplicated or merged.

Regular re-scans may be performed while the affected services are being notified to analyze the remediation behavior of the notified service providers.

\paragraph{Group Assignment}
The recipients are then assigned to the different experimental treatment groups as well as the control group.
This assignment should be stratified by relevant variables (country, industry sector, type of issue (if multiple issues are considered), ...) to ensure homogeneous and comparable groups.

\paragraph{Notification of Recipients}

Immediately before the first notifications are sent, the scan results for the affected services should be checked again to avoid notifying services that have remediated the issue or gone offline in the meantime.
The sending of notifications should be monitored, \eg{}, by collecting reputation information on the sending mail servers.
Depending on the design of the study, it may be necessary to act upon incoming responses and error reports in a timely manner.
When repeated notifications are planned, the dataset should be updated with information from the responses, \eg{}, to reflect changed points of contact.

If self-service tools are part of the study design, their use may be monitored to analyze the behavior of affected service providers.
While analyzing the activities with respect to individual persons may be desirable, such practices are governed by data protection regulations (cf. Sect.~\ref{sec:legal}).

\paragraph{Debriefing}
Especially when the study design involves elements of deception, research ethics (cf. Sect.~\ref{sec:ethics}) mandate that researchers debrief all participants at the end of an experiment. Debriefing can be combined with an invitation to participate in a survey or interview. Members of the control group should also be informed about the vulnerabilities. 

\paragraph{Publishing}

Besides presenting aggregated results in a publication, we encourage researchers to publish those parts of their dataset that help others to check the validity of results, to replicate the experiments, and to reproduce the results independently \cite{acmartifact}, including the source code used for the evaluation. Benureau and Rougier offer practical instructions on achieving reusable code and data \cite{Benureau2017}.
Typically, datasets will have to be carefully sanitized to avoid the unintended release of sensitive information.

\section{Legal and Ethical Obligations}
\label{sec:legalethics}


Researchers not only have to consider ethical problems but also legal issues for their notification studies. In this section, we present different ethical considerations as well as legal questions that occurred in our research.

\subsection{Legal Obligations}
\label{sec:legal}

As legislation differs between countries and the legal situation depends on the specific case, we cannot give concrete legal advice.
Instead, we suggest that researchers discuss their research with legal experts.
To provide starting points for legal consultation, we describe some legal issues that have been discussed during our research.
Specific to the situation in Germany, some advice regarding internet scanning and publication of the scan results can be found in the legal analysis of PrivacyScore \cite{maass2017}.

Notification studies may raise issues in four legal areas:
copyright law, competition law, criminal law, and data protection law.

\paragraph{Copyright Law}

If researchers plan to store the content of a notified party, such as the HTML or JavaScript source code of their website, they might get in conflict with copyright law.
Some legislations have a fair use provision that may allow such storage; others are stricter.
Cooperation with other researchers from less strict legislations might be an option in this case – at least as long as ethical research standards (cf. Sect.~\ref{sec:ethics}) are honored.

\paragraph{Competition Law}

Researchers that intend to release information that can interfere with the competition between entities should take competition law into consideration.
For instance, when we released privacy rankings of several health insurance companies to evaluate their reaction when being notified about these rankings, one health insurance company accused us of violating competition law \cite{MaassWHH19}.

\paragraph{Criminal Law}

When evaluating vulnerabilities, especially when testing whether a vulnerability is exploitable by exploiting it, researchers may come into conflict with criminal law.
For example, in German criminal law, it is forbidden to use default credentials (user: “admin”, password: “admin”) to log into a service without authorization.\footnote{Private communication about § 202a StGB with a public attorney specializing in cybercrime.} While the Budapest Convention on Cybercrime, which has been ratified by 66 states, attempts to harmonize cybercrime legislation ~\cite{CoEBudapest2021}, the local implementation of cybercrime law varies. As a consequence, we refrain from giving more concrete advice.
We stress, however, that researchers typically cannot delegate responsibility for offenses to their institution.

\paragraph{Data Protection Law}

Some countries have strict data protection laws, \eg{}, the GDPR that is implemented within the European Economic Area (EEA).
The GDPR poses restrictions on the processing of personally identifiable information (PII).
PII refers only to individuals and does not address legal entities (Art. 4 No. 1 GDPR); however, notification studies often also address individuals, for example, freelancers.
We note that IP addresses are also considered PII by legal experts\footnote{Gola, DS-GVO, Art. 4 margin number 21 (GDPR legal commentary)}. 
Several duties may arise, such as informing the subjects about the data processing (Art. 13 or 14 GDPR), answering subject data requests (Art. 15 GDPR), allowing subjects to object to the processing (Art. 21 GDPR), or even explicitly asking for their consent (Art. 6(1) a) GDPR).
Also, researchers may have to take technical and organizational measures to protect the data (Art. 24(1) GDPR), such as encryption, storing identifiers to a person separately, or restricting access to specific individuals within their organization.
While the GDPR applies to the EEA, we also note that the GDPR has several opening clauses that allow nation-states to regulate certain aspects differently, \eg{}, for scientific purposes in Art. 89(2) GDPR.
Researchers can ask their data protection officer for consultation.
Within the EEA, public bodies (except courts), including public universities, must appoint a data protection officer (Art. 37(1) a) GDPR).

\subsection{Research Ethics}
\label{sec:ethics}

Besides legal obligations, researchers have to consider the ethical aspects of their work.

Many IRBs focus their attention on whether an experiment classifies as human-subject research. This narrow focus neglects the socio-technical aspects of computer science research. To address ethical aspects more comprehensively, publication venues such as IEEE Security \& Privacy are incorporating mandatory ethics assessments into the reviewing process \cite{hypocrite2021}. Researchers are, therefore, encouraged to deliberate about the ethical implications of their research early on and report them in a systematic fashion within their publications.

One of the cornerstones for ethical research involving information and communication technologies is the \emph{Menlo Report} and its illustrative companion guide \cite{BaileyDKM12}. Accordingly, researchers have to follow four principles, namely (1) Respect for Persons, (2) Beneficence, (3) Justice, and (4) Respect for Law and Public Interest. The Menlo Report can be relied upon in the absence of more concrete guidelines or – as in the study by Dietrich \emph{et al.} \cite{DietrichKBF18} – when no IRB is available for consultation.
However, in practice, its recommendations do not cover all aspects of such studies, and researchers will need to make their own ethical decisions in areas where no broad consensus exists \cite{Vitak2016}.

In the following, we point out selected ethical aspects that arise during notification studies.
Firstly, such studies involve scanning the services of third parties without obtaining their permission. Secondly, researchers interact with humans, \eg{}, end users or employees of affected service providers.

\paragraph{Network Scanning}

For scanning and data collection, researchers should follow best practices developed by the network measurement and security scanning communities. For instance, Durumeric \emph{et al.} \cite{Durumeric2014} describe seven practices for good internet citizenship: (1) coordinate with local administrators to handle inquiries, (2) verify that scans will not overwhelm the upstream network, (3) signal benign nature of scans via web pages and DNS entries, (4) explain purpose and scope of scans in communications, (5) provide a simple means of opting out, (6) conduct scans no longer or more frequent than necessary, and (7) spread scan traffic over time and source addresses.

\paragraph{Human-Subject Research}

Many notification studies classify as human-subject research.
This is especially true when recipients are deceived, for instance, when researchers do not disclose the fact that they run a study that analyzes the behavior of recipients.

There is an ongoing debate about the ethical obligations for internet-based human-subject research \cite{Vitak2016}. The details are beyond the scope of this paper. Bravo-Lillo \emph{et al.} provide some practical guidance \cite{Bravo-Lillo2013a}.
In notification studies, researchers will typically have to debrief all study participants – including the control group – at the end of the experiment. Moreover, the study design should allow participants to opt-out of scanning and further notifications.

Researchers should familiarize themselves with the processes for ethical review and seek approval as early as possible to avoid delays caused by missing ethics approval.

\section{Data Collection}
\label{sec:datacollection}


To collect a dataset of operators to notify, notification studies usually have an initial data collection phase in which large-scale scans of the internet are used to detect machines suffering from the issue at hand.
In this paper, we do not discuss the source of the list of systems to be scanned.
Some studies scan the entire IPv4 address space, while others rely on lists of domains.
We note that the most common source of domains, the Alexa Top Million, has been criticized as unstable and potentially unrepresentative \cite{Scheitle2018,Pochat2019}.
The \emph{Tranco list} has been proposed as an alternative \cite{Pochat2019}.

In this section, we begin by considering best practices for developing and operating scanning infrastructure. We then discuss the design of the scanning infrastructure, including what data should be collected, and close with a recommendation on scheduling the periodic scans that are a core part of many notification studies.

\subsection{Infrastructure}
Notification studies span several months and frequently necessitate changes to the scanners and other infrastructure over time as the software is extended, bugs are fixed, and capacity problems are revealed.
This makes automation critical to ensure a consistent and efficient operation of the system.

\paragraph{Use Version Control and Issue Tracking}
All developed software should be tracked in a version control system like Git. This allows a simpler collaborative development process and is also critical for replicability and reproducibility \cite{Benureau2017}. Ideally, this should be combined with an issue tracker where planned features, bugs, and other details of the software development can be documented. 

\paragraph{Automate Deployments}
After the software is written, it must be deployed on the production systems. For all non-trivial systems, we strongly recommend using an automated deployment process using tools like Ansible or Docker. Having these tools pull the data directly from the version control system also disincentivizes the antipattern of making manual changes to deployed infrastructure that are not tracked in version control and makes documenting the exact version of the software that created a result easier \cite{Benureau2017}.

\paragraph{Operate Test Systems}
Having self-operated test systems where the expected result of a scan is known helps to test the detection software during development.
During regular scans, test systems can verify that the detection software and test harness are still operating as expected.
Also, these systems can serve as a \emph{dead man's switch}, \ie{}, when not being scanned during the expected interval, they notify the researchers.
Services for a dead man's switch reporting include Healthchecks.io, Dead Man's Snitch (deadmanssnitch.com), or PushMon.com.
Some of these provide a free tier that is likely to be sufficient for a notification study.

\paragraph{Monitor the Infrastructure}
Software and machines can fail in surprising ways, especially in situations of high load. The systems should thus be monitored using the aforementioned test cases, checking for timeouts, monitoring for exceptions or implausible results, and checking the utilization of resources (RAM, disk space). The monitoring should be combined with an alerting mechanism to inform the operators about errors as they occur. 

\paragraph{Make Backups}
Scanning infrastructure, like any other computer, can suffer from data loss, either through hardware failure or software issues \cite{Cui2011}. The data should thus be backed up through regular automated backups to \emph{at least} one (and preferably multiple) other machine(s), ideally on a completely different network and physical location. This recommendation is particularly important if the scanner infrastructure is hosted on third-party infrastructure outside the direct control of the researchers (\eg{}, AWS), where it may be disabled without consulting the researchers if the company receives abuse notifications.

\subsection{Developing the Scanner}
A notification study typically uses two kinds of software: a detection software that scans for the issue being reported as well as a test harness that executes the scans on all previously collected targets and stores the results.


\paragraph{Choose the Right Tool}
Depending on the type of issue, the detection software can range from a simple script to a fully instrumented browser.
For scanning websites using browser instrumentation, OpenWPM \cite{Englehardt2016} and privacyscanner \cite{privacyscanner} are two pieces of software that can be used or extended.
OpenWPM instruments Firefox, while privacyscanner instruments Chrome.
While browser instrumentation represents the reality more accurately, \eg{}, being able to detect dynamic content that an HTTP library does not see, it is also more complex and resource-intensive.
Researchers should thoroughly evaluate the limitations of their detection software.

\paragraph{Web-specific: Expect Cookie Banners and Bot Detection}
Since the coming-into-effect of the GDPR, many websites are using cookie consent banners that may hide specific parts of the website until consent for tracking has been given.
Similarly, some websites attempt to block access for automated programs using bot detection software.
If the issue in question depends on measuring the presence of or interactions with a specific third-party service, this may lead to false-negative results.
Researchers should be aware of this risk when planning their study.

\paragraph{Web-specific: Decide how to Handle Redirects}
Websites necessitate a number of special considerations due to their dynamic nature.
One of them is the existence of redirects between different domains.
This can cause two classes of issues: firstly, two domains referring to the same final domain (making two seemingly distinct scans return identical results and thus biasing the dataset), and secondly, one domain changing which final domain it refers to (making two scans of the same domain return results for different websites over the course of the study).

These redirects have multiple implications for notification studies.
Firstly, the scanners need to support forwards, which can also be triggered through JavaScript and thus invisible to simple downloading scripts that only follow HTTP redirects.
Secondly, if redirects are followed, the researchers need to decide if they follow the redirect every time, or follow it once, save the final URL, and then scan this final URL for all future scans.
The first approach mirrors the behavior of users, while the latter leads to more consistent results.
Finally, regardless of which strategy is chosen, the scanning system should save the URL after following all redirects as part of the results to facilitate later analysis.

If redirects are followed, researchers also need to consider how to handle results obtained from intermediate pages.
For example, if the use of a specific third-party service is of interest, what happens if it is only used on an intermediate website that then forwards to a different website?
The scanners need to ensure that any saved results can be correctly attributed to the intermediate pages to facilitate a later exclusion if this is desired.

If two or more domains forward to the same final domain, the operator of that domain may gain undue influence on the evaluation, as remediation by this one operator may be counted for more than one website.
These cases need to be considered in the evaluation and addressed.

\paragraph{Collect Enough Data}
It is not always possible to know in advance which data will be needed for the evaluation.
We thus recommend collecting as much (meta)data as possible, both about the scan target and about the machine running the scan.
Potentially relevant data points include: which machine was running the scan, the version of all software and libraries in use by the scanner, which IP address was scanned (if scanning based on DNS names), HTTP response codes (for websites), the raw output of any external scanning tool, and extensive log files with timestamps.
For a list of metadata recommended for replicability, see Benureau and Rougier \cite{Benureau2017}.

As previously discussed, however, the data collection also needs to consider legal obligations and ethical aspects.
Care must be taken not to impact the operation of the target server.
Finally, in some cases, storage space and network throughput may be of concern.

\paragraph{Web-specific: Archive Websites}
Some situations necessitate verifying the state of a website at a specific point in time, for example, when implausible results from past scans should be validated.
In these cases, it can be valuable to have an archived version of the website to refer back to.
Such archives can either be created using the Wayback Machine of the Internet Archive\footnote{The Wayback Machine can be instructed to create a snapshot of a website by sending a GET request to \texttt{https://web.archive.org/save/[website-url]}. When doing this at scale, researchers should identify themselves by setting a user-agent with their contact information in case of problems.} or using tools like webrecorder (github.com/webrecorder).

\paragraph{Plan for Different Types of Scans}
As described in Sect. \ref{sec:timeline}, notification studies may require initial scans and regular re-scans.
Initial scans use a larger dataset to find targets, while regular re-scans only address previously found targets.
Thus, the test harness should support changing the dataset as well as one-off and regular scans.




\paragraph{Identify the Scanner}
When not covertly scanning, researchers should identify their scanning software or hosts to reduce abuse reports from recipients who would otherwise misinterpret the scan as an attack.
To identify a scanning host, researchers can set an appropriate PTR DNS record (\eg{}, leak-study.yourinstitution.org), or host a website on the IP address of the scanner \cite{Durumeric2013}.
For web-based scans, the User-Agent header can identify the scanner and provide a link for more information.

\paragraph{Know the Error Classes}
Detection software has to deal with various error conditions, possibly supported by the test harness.
Researchers, therefore, should evaluate the behavior of their software in such error conditions.
These conditions include failure of the network connection in the middle of a test, an unreachable target, or a test host that ran out of disk space or memory.
Also, the detection software or any software it calls might return an error or freeze.
A test harness can deal with many cases, \eg{}, by implementing a retry mechanism if the detection software fails,
or killing the detection software if it runs an unusually long time.
Moreover, when running detection software in parallel, some state might be unintentionally shared, resulting in errors or erroneous data.



\paragraph{Prepare for Unknown Errors}
Scanning on a large scale will likely produce new error classes.
Thus, researchers should develop their software to detect deviating behavior and report those cases.
Deviating behavior includes unexpected output, an unusual scan duration (\eg{}, due to deadlocks/livelocks or rate limiting), unlikely changes in measured values (\eg{}, response size drops from many KiB to a few bytes), or uncaught exceptions.
To report deviating behavior, we recommend \emph{Sentry} (\url{https://sentry.io}), which centrally logs uncaught exceptions (including stack traces) and any other information researchers wish to get reported.

\subsection{Scanner Operation}
\label{sec:lessons:scanning:collection}

Once the scanner has been developed, it needs to be put into operation to perform regular scans.
We give recommendations on the scanning schedule and infrastructure deployment.

\paragraph{Inform the Network Operator}
Researchers should contact their network operator beforehand and explain the scanning engine.
Otherwise, the network operator might be surprised by unusual traffic patterns and considers them a threat, even without incoming abuse reports.
Also, researchers should ask their network operator whether any technology is in place that could interfere with the scan, such as firewalls, IDS appliances, or connection throttling.

\paragraph{Begin Early}
After the first scans determine the list of systems that will be included in the study, researchers will usually plan for periodic scans to update the current remediation status of all included systems.
We recommend beginning these scans as early as possible and, if feasible, at least 1–2 weeks before the first notifications are sent.
This serves two purposes: it tests the scanning infrastructure in action and validates that no unexpected problems occur, and it collects a dataset of system behavior before any outside intervention.
This can be used to validate that the different experimental groups show similar behavior before the intervention, increasing the confidence that any observed differences are due to the notification (and not fundamental differences between the groups).
If the groups already diverge within this timeframe, the group allocation strategy should be reconsidered.

\paragraph{Scan Often}
Regular scans should be executed often, ideally several times per day for each target for several reasons.
Firstly, scanning often allows for a more detailed and fine-grained analysis of potential remediations.
Secondly, more data points allow for better interpolation of missing or erroneous scan results, \eg{}, due to connection issues.
Finally, some systems may behave differently depending on the time, \eg{}, a website that has a day and a night version.

\paragraph{Scan From Multiple Places}
We recommend running several redundant copies of the infrastructure on different machines and networks to prevent losing information when one machine or network fails.
In addition, a comparison of data from different copies may find additional errors or edge cases in the scanning stack.

\section{Interacting with Contacts} \label{sec:interaction}
Notification studies necessarily involve communicating with large numbers of system operators.
In this section, we consider different aspects of sending notifications to the operators and handling their responses.
We also discuss how a self-service tool for recipients can reduce the burden on the researchers and give advice on combining the study with a survey.

\subsection{Sending Notifications}
After a list of affected systems has been found, the next step is to identify the relevant point of contact and sent the notifications.
The choice of contact method is diverse and often a central point of notification studies, so we will not go into detail on the collection of address information itself.
Instead, we assume that a set of addresses is known and proceed from there.

\paragraph{Deduplicate the Contacts}
A single operator may be responsible for more than one affected system.
This can have multiple implications: firstly, the operator may have an outsized impact on the overall result if they operate a large number of affected systems, as they are likely to remediate all (or none) of their systems at once.
Secondly, if multiple experimental groups exist, one operator can be part of more than one group and receive multiple notifications, which may confound any analysis of the effectiveness of individual groups.
To address this, operators should be deduplicated and grouped on a best-effort basis.
This is easiest when using manual data collection and impossible when using standard email aliases (RFC 2142), which derive the contact address from the scanned domain without consulting any external database of contact information.
Researchers should be aware of the effects this (lack of) grouping can have on their evaluation.

\paragraph{Automate Message Generation and Sending}
When dealing with many recipients, manual work is prone to errors.
We recommend writing scripts that generate the text for all recipients, especially when recipients are split into groups with different treatments.
For letters, scripts can generate \LaTeX{} source code that is compiled into PDF files for printing.
When not sending the emails via a script directly, researchers can use the Thunderbird plugin
\emph{Mail Merge}\footnote{See \url{https://addons.thunderbird.net/addon/mail-merge}.}.
This plugin sends emails according to a template for which it reads template variables and recipients from a CSV file.

\paragraph{Run Sending Tests Beforehand}
To catch mistakes beforehand, researchers should send all their notification emails via a test mail server that does not deliver the messages to the recipients but allow researchers to view the sent mails. 
MailHog (github.com/mailhog) and MailSlurper (mailslurper.com) are two examples of software that present the sent mails in a web-based frontend.

\paragraph{Implement SPF and DKIM}
To authenticate senders and reject spam, mail servers rely on DKIM and SPF DNS records.
Receiving mail servers may reject notifications if sending servers do not implement DKIM or SPF properly.
We found that even large universities can fail to implement SPF and DKIM \cite{usenix2021}, so researchers should always validate the servers' configuration.

\paragraph{Check for Reputation}
If researchers operate their own mail server, they should subscribe to reputation monitoring systems such as
Microsoft Junk Mail Reporting Program
or Google's Postmaster Tools\footnote{See \url{https://mail.live.com/mail/services.aspx} and \url{https://postmaster.google.com/}}.
Furthermore, researchers should check spam blocking lists regularly, such as SpamCop.net and Spamhaus.org.


\paragraph{Stretch Sending of Emails}
Sending bulk emails may result in hitting a mail server's rate limit.
We recommend that researchers test the desired rate beforehand to avoid unexpected errors.
Furthermore, bulk emails may trigger spam filters that look for emails with similar content from the same sending mail server.
We had good experiences with sending an email every thirty seconds.

\paragraph{Prepare for Undeliverable Emails}
Emails to some recipients might be undeliverable.
To account for undeliverable mails during analysis, researchers should check and keep track of \emph{bounces}.
Undeliverable emails and bounces manifest in different forms, which makes this task challenging.

Firstly, there are delivery delays due to retries.
Mail servers may retry delivering an email for several days, with some mail servers reporting the retry, while others only inform the sender when giving up.
However, there is no guarantee that the sender will be informed at all.
Thus, researchers should not make assumptions about delivery success right after sending a message.

Secondly, bounces are not standardized.
Receiving mail servers might reject an email with an error code right away when it is delivered via SMTP.
Some errors are permanent (\eg{}, recipient not known), others are temporary (\eg{}, quota is full) and may result in delivery retries.
In some cases, the sending mail server informs a user about the rejection during the submission process; in other cases, sending servers inform users about delivery issues with an email later.
Also, the \emph{receiving mail server} might inform the sender about the undeliverable email.
These delivery failure notices may or may not contain information about the email that was undeliverable.
Senders may not receive such notice at all.

Finally, some delivery failure notices are not bounces, but normal emails that were sent as an auto-reply to an incoming message, telling the researcher that the message was not read by a human.
In the end, it is impossible to automatically handle all cases. Manual work is required to classify automated delivery status notifications and assign them to the correct recipient.

\subsection{Handling Responses}
Once they have received the notification, system operators may initially distrust it or have questions about the details.
In these cases, they will frequently seek to get into contact with the sender of the message.
How the researchers react to their questions can have a large impact on their behavior and thus on the results of the notification campaign.
We thus highlight a number of experiences and best practices for handling responses.


\paragraph{Build a Frontend}
Answering a recipient's response often requires information about them.
Relevant pieces of information include the reported issue (if different issues are reported) or the group the recipient is assigned to (if researchers vary the sent notification).
A custom tool can help to find those pieces of information efficiently, e.\,g., while answering a phone call.
Useful features include re-scanning the target to get the current state of the issue and a fuzzy search to find the recipient or their website.
When researchers plan to analyze the responses in more detail, \eg{}, coding them, a tool can help keep the responses organized.


\paragraph{Expect Deviating Communication Channels}
While researchers provide their contact information to the notified recipients, they should expect to receive responses to unrelated (even private) email addresses and via other communication channels such as phone and social networks.
We found that some notified recipients used search engines to validate the legitimacy of the sender and find communication channels they preferred, such as a phone number \cite{usenix2021,maass2021leaks}.
Researchers should consider proactively including a phone number and preferred timeframe for calls in the contact information.

\paragraph{Expect Colleagues and Others to be Contacted}
As described above, notified recipients will search for alternative contact information.
Consequently, some recipients will find other contact information such as those of secretaries or the central phone number of the institution and use those to establish contact.
Recipients may also choose to complain at higher hierarchy levels, such as the dean.
We, therefore, recommend informing the respective persons beforehand about the study to avoid surprises.
It might be useful to prepare a one-page document describing the notification study and whom to contact or forward to.

\paragraph{Know how to Handle Gifts}
Many recipients are grateful for the help; some also offer gifts or payment.
Accepting gifts or payment may have negative consequences. 
Depending on the legislation, when working for a public body, accepting gifts may represent a criminal offense and can lead to termination of the work contract.
Private companies may have internal compliance rules that forbid accepting gifts.
Since some recipients send gifts without asking first, we recommend that researchers discuss how to handle these cases with their institution.
Researchers that cannot keep gifts could approach non-profit organizations and ask whether they accept donations of gifts and attest the receipt.

%
\paragraph{Prepare for Misunderstandings and Threats}
Some recipients misunderstand the message as spam, scam, legal threat, or defamatory.
This fact can lead to uncomfortable messages or phone calls that include legal threats such as sending a cease-and-desist letter or suing researchers \cite{MaassWHH19,usenix2021,Cetin2017}.
Again, we recommend consulting legal experts beforehand; see Sect. \ref{sec:legal} for details.
Researchers should offer those recipients to exclude them from further messages.

\paragraph{Have a Help Policy}
Helping recipients in remediating an issue may influence the results, e.\,g, when remediation rates are measured.
Not helping, however, also has an influence since it may lead to resentment.
In addition, some recipients may falsely claim that the reported issue has been remediated.
Again, telling recipients of their false assumption may distort the experiment.
Researchers should decide on a help policy as part of their study design.
Moreover, ethical considerations must be taken into account when not helping,
especially if the reported issue may impose harm on others.



\paragraph{Expect Unrelated Requests}
Helping recipients may result in further requests for help with unrelated problems.
Similarly, when dealing with a compliance issue, recipients may ask for legal advice.
Note that some legislations have restrictions on giving legal advice, \eg{}, in Germany the \emph{Rechtsdienstleistungsgesetz} restricts non-lawyers in giving advice.
We recommend politely declining to help in all such cases.


\subsection{Self-service Tool}
Typically, recipients want to know whether their remediation attempt was successful.
Instead of answering that question individually, researchers may provide a tool for this purpose.
In the following, we discuss various design decisions.

%

\paragraph{Provide Clear Instructions}
Users of the tool should be able to understand the tool's purpose and how to use it.
Instead of only showing results, the tool should help the user to interpret the results to avoid unnecessary support requests.
If possible, the tool should provide extensive information on how to remediate the issue, ideally with code or configuration examples.
Note, however, that clear instructions might not be sufficient, \eg{}, a study by \c{C}etin \emph{et al.} reported recipients that had trouble understanding or using a tool correctly despite clear instructions on the tool page \cite[p. 7]{Cetin2017}.
Thus, if possible, the tool should also detect common forms of incorrect usage and provide specific guidance in these cases.


\paragraph{Do not Restrict the Targets}
Some recipients may operate several websites or hosts and want to scan all their systems for the reported issue.
Thus, researchers should consider allowing scanning of arbitrary targets.
In addition to reducing support requests asking the researchers to scan additional targets, allowing arbitrary targets can provide additional information on the remediation behavior of the recipients.
Offering a public and unrestricted tool, however, is subject to ethical considerations if the tool can be abused or harm others.


\paragraph{Publicize the Tool}
Recipients might distrust the link to the tool in the notification, expecting it to be a scam.
To check whether the link is legitimate, some recipients may attempt to find the tool using a search engine.
Thus, assuming the tool does not rely on personalized links, researchers should submit the tool's website to the index of search engines, making sure it can be found.
In addition, linking the tool from a university website (if this is compatible with the study design) and giving it a (semi-)professional look can increase the trust in the tool.


\paragraph{Collect Tool Usage Data}
Scans performed by the tool should be logged. 
This includes the scan time, the target, and results.
Collecting this data can help to answer various questions, for example:
How often was the target scanned before the issue was remediated?
How much time passed between the first and last scan?
Does the state of the issue change on scans, \ie{}, did the user make mistakes when trying to remediate the issue?
Were other sites scanned? If yes, which sites? 
Do they follow the same remediation pattern?
We strongly recommend that researchers deploy an additional \emph{internal instance of the tool} so that they can scan targets, having the same view as regular users, without polluting the data collection of the regular tool instance.


\paragraph{Be Aware of Alternative Tools}
For some issues, there might be more than one check tool.
Some recipients might try other tools to get information, possibly conflicting with the information the researchers' tool provides.
This conflicting information might confuse recipients, leading to additional questions and support requests.
Researchers should therefore look for other tools and make themselves familiar with those, especially with their limitations or errors.



\subsection{Survey}
While interactions with recipients can be a source of qualitative data, their free-form nature does not lend itself to answering quantitative questions.
Quantitative questions can be addressed with a survey.

\paragraph{Decide When to Run the Survey}
The first question when planning a survey is at which point in the process it should be sent out.
Some prior studies sent the survey together with the notification \cite{Li2016Usenix,Cetin2017,Zeng2019}, while others sent it later in the process \cite{usenix2021,Durumeric2014,Stock2018} (\eg{}, with a debriefing message).
This decision may, in some cases, be dictated by the experimental setup (\eg{}, if the fact that the messages are sent as part of a study should initially be hidden from the recipients, including a link to a survey with the notification message will usually not be possible).
On the other hand, sending the survey weeks or months after the notification may mean that some recipients will not be able to remember their initial perception of the notification message.

\paragraph{Distinguish the Groups}
If the notification study employs multiple experimental groups, researchers should ensure that these groups' answers can be distinguished in the survey results to allow group-specific evaluations of the responses. This requirement can be addressed in a privacy-preserving way by using different instances of a survey and sending group-specific links to recipients. Using the more anonymous group-specific links instead of personalized links with unique IDs \emph{for each recipient} might encourage recipients to participate.

\paragraph{Consider Surveying the Control Group}
Since the control group needs to be notified about being part of the study for ethical reasons, this can be a good opportunity to send them a survey as well.
While they cannot provide insight into aspects of the notification itself, they may be able to provide further data on the sources of the vulnerability or misconfiguration in question, and gain further insight into the perspective of operators.

\section{Publication of Results} \label{sec:publication}
After data collection and analyses are completed, the results will usually be published in a scientific venue.
To ensure reproducibility, \emph{data} and \emph{source code} should be published whenever possible. At the same time, in order to avoid putting systems and operators at risk, researchers need to ensure that they do not release information that would allow others to infer who was part of the study and what their results were. While evaluation code can usually be released without worrying about deanonymizing study participants, releasing the dataset is more difficult.

\paragraph{Ensure Reproducibility}
The code used for evaluation should follow best practices for reproducibility. A comprehensive source of best practices for code and documentation has been published by Benureau and Rougier \cite{Benureau2017}.

\paragraph{Sanitize the Data}

Before the public release of any dataset, all records need to be sanitized. An initial – typically not sufficient – step is the removal of all identifiers that may allow others to infer the identity of services or operators.
This includes obvious aspects like IP addresses and domain names, but also unique identifiers that can be linked to a server, such as TLS certificates, identifiers in the network traffic, cookies, etc.

When it is not possible to remove the identifiers altogether (\eg{}, because they are needed to separate different servers that are using the same identifying information), identifiers can be \emph{pseudonymized}, i.\,e., replaced with a unique number or string. The pseudonym should not be directly derived from the original identifier (\eg{}, using an unsalted hash function), as this may allow others to re-identify records. Pseudonymization best practices are beyond the scope of this paper. A comprehensive guide has been published by ENISA \cite{pseudoenisa}.

\section{Conclusion}

Large-scale vulnerability notifications are an important building block in improving security and privacy on the internet – and the search for the most effective set of parameters is still ongoing.
Notification studies are complex experiments with both technical and interpersonal challenges. The design space of such studies is limited by legal and ethical obligations.
Drawing from experience gained during several such studies, we presented best practices for data collection, message delivery, interaction, and tool support as well as the integration of surveys and considerations for publication.

While compiling this collection, we were reminded of two overarching lessons that we learned during our studies. Firstly, using tools to automate data handling turned out to be a life-saver, both for our workload as well as data quality. Secondly, expect the unexpected: time spent in the planning phase to account for potential failure cases (and methodological weaknesses) pays off at the end.

The best practices documented in this paper can inform future notification studies to avoid common pitfalls and maximize benefit for all involved parties, thus, ultimately helping to improve security and privacy on the internet.

\section*{Acknowledgements}

This work has been co-funded by the DFG as part of project C.1 within the RTG 2050 “Privacy and Trust for Mobile Users” and by the German BMBF and the Hessen State Ministry for Higher Education, Research and the Arts within their joint support of the National Research Center for Applied Cybersecurity ATHENE.

\bibliographystyle{ACM-Reference-Format}
\bibliography{bibliography}

\end{document}